\documentclass[a4paper, 10 pt, conference]{ieeeconf} 

\IEEEoverridecommandlockouts                              % This command is only needed if 
\pdfoutput=1

\usepackage{amsmath}
\usepackage{graphicx}
\usepackage[caption=false]{subfig}

\usepackage{amssymb}
\usepackage{mathtools}

\newcommand{\be}{\begin{equation}}
\newcommand{\ee}{\end{equation}}

\newcommand{\bee}{\begin{equation*}}
\newcommand{\eee}{\end{equation*}}

\begin{document}
\title{Modelling arterial travel time distribution using copulas}
\author{Adam Samara, Felix Rempe and Simone Göttlich% <-this % stops a space
\thanks{Adam Samara and Felix Rempe are with BMW Group, Petuelring 130, 80788, Munich, Germany      {\tt\small adam.samara@bmw.de}, {\tt\small felix.rempe@bmw.de} }%
\thanks{Adam Samara and Simone Göttlich are with the Department of Mathematics, University of Mannheim, 68131 Mannheim, Germany
        {\tt\small asamara@mail.uni-mannheim.de}, {\tt\small goettlich@uni-mannheim.de}}%     
}        

% The paper headers
\markboth{Journal of \LaTeX\ Class Files,~Vol.~6, No.~1, January~2007}%
{Shell \MakeLowercase{\textit{et al.}}: Bare Demo of IEEEtran.cls for Journals}

\maketitle
\thispagestyle{empty}

\begin{abstract}
%\boldmath
The estimation of travel time distribution (TTD) is critical for reliable route guidance and provides theoretical bases and technical support for advanced traffic management and control. The state-of-the art procedure for estimating arterial TTD commonly assumes that the path travel time follows a certain distribution without considering segment correlation. However, this approach is usually unrealistic as travel times on successive segments may be dependent. In this study, copula functions are used to model arterial TTD as copulas are able to incorporate for segment correlation. First, segment correlation is empirically investigated using day-to-day GPS data provided by BMW Group for one major urban arterial in Munich, Germany. Segment TTDs are estimated using a finite Gaussian Mixture Model (GMM). Next, several copula models are introduced, namely Gaussian, Student-t, Clayton, and Gumbel, to model the dependent structure between segment TTDs. The parameters of each copula model are obtained by Maximum Log Likelihood Estimation. Then, path TTDs comprised of consecutive segment TTDs are estimated based on the copula models. 
The scalability of the model is evaluated by investigating the performance for an increasing number of aggregated links.
The best fitting copula is determined in terms of goodness-of-fit test. The results demonstrate the advantage of the proposed copula model for an increasing number of aggregated segments, compared to the convolution without incorporating segment correlations.

\end{abstract}

%\IEEEpeerreviewmaketitle

\section{Introduction}

Travel time reliability (TTR) estimates are of major importance for travelers and transport managers as they are very informative for decision making and planning schedules. 
TTR has been increasingly recognized as an important measure for estimating the operation efficiency of road facilities, assessing alternative management strategies \cite{SHRP1}, and providing travelers with route guidance \cite{SHRP2}.
TTR is defined as the consistency or dependability in travel times, as measured from day to day and/or across different times of the day \cite{lyman}.
In \cite{bates} it is suggested that the analysis of TTR is as important, if not more important than, the traditional analysis of average travel time. In order to fully assess TTR, travel time distribution (TTD) needs to be determined as a prior. This makes it possible to measure the risk for on-time arrival probability and find a path for risk-averse travellers \cite{zeng}.

A common approach is to aggregate segment TTD to a joint distribution by assuming independence between individual segment TTD \cite{dailey}, \cite{itsc19}. However, such an approach appears inaccurate since travel times on successive segments are essentially dependent. For example, when one segment becomes congested, the neighbouring segment also gets affected by this congestion.

This work addresses this problem, by adopting the copula model in econometrics \cite{trivedi} to assess path TTD given a segment set by accounting for correlation between segment travel times. Using copulas for estimating travel time was proposed by \cite{CHEN2017} and \cite{CHEN2018}. The proposed methodology was evaluated by \cite{CHEN2017} for the through movement of two arterials in Shanghai, China and Los Angeles, California, based on automatic vehicle identification (AVI) and Next Generation Simulation (NGSIM) data, respectively, while
the copula model was evaluated by \cite{CHEN2018}  by utilizing VISSIM simulation with calibration to generate travel time data on one arterial in Hangzhou, China.
The comparison in both studies was made for path TTDs estimated by the copula model, the convolution and the empirical distribution fitting approach, indicating a superior performance of the copula model. 
Table \ref{tabshortcomings} gives an overview of both studies.
\begin{table}[h]
\label{tabshortcomings}
\caption{ Related work using copula models.}
\begin{center}
\begin{tabular}{lll}
\hline
          & Chen et al. \cite{CHEN2018}  & Chen et al. \cite{CHEN2017}  \\
\hline 
Data   & VISSIM   & AVI and NGSIM   \\

Numb. of segments  &  Two   &   Three   \\

Copulas          &    Gaussian       &    Gaussian, FGM, AMH, Frank   \\
\hline
\end{tabular}
\end{center}
\end{table}

However, the data used by both studies does not represent day-to-day travel time observations. In addition, only two and three segments are aggregated, respectively.
This leaves the performance of the proposed copula model for an increasing number of segments an open question, as a path is comprised of several segments when using route guidance systems. 

This work shows that the copula model provides a better solution than the traditional convolution model even when using real day-to-day GPS data collected over the period of one year, and for an increasing number of segments. In addition several copulas are compared, and the best fitting copula is determined.

One distinct feature of the copula model when compared to multivariate distributions is that the dependence structure is unaffected by the types of marginal distributions, which enables greater flexibility in correlating individual segment TTD.
For this study, path TTD for a major urban arterial in Munich, Germany, is investigated.

This article is structured as follows: In the first section, the copula theory as well as the estimation procedure for copula models is described. Then, a case study is conducted for the study site using historical travel time data provided by BMW Group.
Path TTDs are estimated by copula models, first by aggregating two segment TTDs, then by aggregating ten segment TTDs. These results are compared with the estimates obtained without considering correlations between segment travel times and empirical distributions. The last section draws a conclusion and gives an outlook for future work.

\section{Methodology}
In this section a brief overview of the underlying theory of copulas adopted from \cite{roger} is given. In addition the proposed copula model is described, which was implemented based on \cite{yan}, \cite{kojadinovic} and \cite{hofert}.

\subsection{Mathematical preliminaries}
A  road network can be represented as a directed graph
$\mathcal{G} = (\mathcal{V}, \mathcal{E})$
which is an ordered pair of a (finite) set of vertices $\mathcal{V}$ and a set of edges $ \mathcal{E}$, representing geolocations and road links connecting these locations, respectively. An edge $e \in \mathcal{E}$ comprises a pair of two vertices $v_1, v_2 \in \mathcal{V}$. 
Besides we have $o, d \in \mathcal{V}$ , representing the origin and the destination, respectively. 
The travel time for each link is represented as a random variable $x$, which is derived from historical data. Thus, empirical link TTD is discrete. For characterizing link TTD as continuous probability density function $f(x)$ with distribution function $F(x)$, both parametric and nonparametric estimators can be used.
Examples for parametric estimators are Normal, Lognormal, Gamma, and Weibull, while Kernel Density Estimation and Gaussian Mixture Model are examples for nonparametric estimators.

 A path from $o$ to $d$ is comprised of several successive links. As historical data for entire paths are not available, path TTD is obtained by aggregating link TTD, which is explained below.

\subsection{Copulas}
Copulas are functions that relate multivariate distribution functions of random variables to their one-dimensional marginal distribution functions.
 According to Sklar’s theorem \cite{sklar1996}, for an $n$-variate distribution function $F (x_1, ..., x_n)$ with marginal distribution functions $F_1(x_1), ..., F_n(x_n)$, there exists a certain copula function $C$ which meets the relationship
\be
\label{eq2106}
F (x_1, ..., x_n) = C (F_1(x_1), ..., F_n(x_n)).
\ee
If marginal distributions are all continuous, $C$ is unique.
Based on Sklar's theorem the concept of copula provides an efficient way of modeling dependent variables.
Following from Sklar's theorem the joint distribution of $f(x_1, ..., x_n)$ can be obtained by 
\begin{equation}
\label{eqsklardens}
f (x_1, ..., x_n) = c (F_1(x_1), ..., F_n(x_n))  \prod^{n}_{i=1} f_i (x_i)
\end{equation}
with copula density
\be
c (F_1(x_1), ..., F_n(x_n)) = \frac{\partial C (F_1(x_1), ..., F_n(x_n))}{\partial F_1(x_1),  ..., \partial F_n(x_n)}
\ee
and marginal distribution functions $f_i (x_i)$.

There exist two major families of copulas: Archimedian and elliptical copulas.
Archimedean copulas are very popular because they are easily derived and they are capable of capturing wide ranges of dependence. The definition of the Archimedean copula is based on the generator function $\varphi$. Archimedean copulas take the form
\be
C (u_1, ..., u_n) = \varphi^{(-1)} (\varphi(u_1) + ... + \varphi(u_n)),
\ee
where $\varphi^{(-1)}$ is the pseudo-inverse of $\varphi$.
 The reason for Archimedean copulas' popularity in empirical applications is that it produces wide ranges
of dependence properties for different choices of the generator function. 
Two of the most frequently used archimedian copulas are the Clayton Copula with the generator function 
\begin{equation}
\varphi^{\text{Clayton}}(u) = (1 + u)^{(-1 /  \alpha) }, 
\end{equation}
and the Gumbel Copula with the generator function 
\begin{equation}
\varphi^{\text{Gumbel}}(u) = \exp(-u^{(-1 /  \alpha)}), 
\end{equation}
where $u$ is the marginal distribution and $\alpha$ is the Copula parameter, which describes the dependency between the random variables $x_i$. It can be determined by rank correlation coefficient, e.g, Kendall correlation coefficient \cite{roger}. 
A Clayton copula is able to capture lower tail dependence, and a Gumbel copula is able to capture upper tail dependence. 

The elliptical copulas differ from the Archimedean classes of copulas in the approach that only an implicit analytical expression is available. These copulas are derived from the related elliptical distribution.
The first example of elliptical copula is the Gaussian copula, which belongs to normal distribution, defined as
\begin{equation}
 C^{\text{Gauss}}_{R}(u_{1}, ..., u_{n}) = \Phi_{R} (\Phi^{-1}(u_{1}), ...,\Phi^{-1} (u_{n})),
 \end{equation}
 where $\Phi_{R}$ is the joint normal distribution with correlation matrix $R$ and $\Phi^{-1}$ is the quantile function of the univariate standard normal distribution. For the purpose of this work a uniform correlation structure was used for the correlation matrix, so that $R = (1-p)I_\rho + \rho \mathbb{I} \mathbb{I}^{\prime}$ with correlation coefficent $\rho$.
The second example is the Student-t copula, which belongs to the t-distribution, defined as
 \begin{equation}
 C^{\text{Student t}}_{\nu,R}(u_{1}, ..., u_{n}) = t_{\nu, R} (t_{\nu}^{-1}(u_{1}), ...,t_{\nu}^{-1} (u_{n}) )
 \end{equation}
 where $t_{\nu, R}$ is the joint Student distribution with correlation matrix $R$ and $\nu$ degrees of freedom, and $t_{\nu}^{-1}$ is the quantile function of the univariate Student distribution. In this case $R$ was chosen to be uniform. Unlike the Gaussian copula the Student-t copula is able to capture tail dependence.
  
\subsection{Estimation of copula models}
A copula model is estimated in two steps. First the segment TTDs are estimated from empirical GPS data, and then the copula parameters are fitted.  
In this study finite Gaussian Mixture Model (GMM) \cite{rasmussen} was used, as it showed an accurate fit.
The finite GMM with $k$ components is represented as 
\be
p (y | \mu_1, ..., \mu_k; s_1, ..., s_k; \pi_1, ..., \pi_k) = \sum^{k}_{j=1} \pi_j \mathcal{N} (\mu_j, s_{j}^{-1}),
\ee
where $\mu_j$ are the means, $s_j$ are the inverse variances, $\pi_j$ are the mixture weights and $\mathcal{N}$ is a normalized Gaussian with specified mean and variance.
The parameters of the GMM are obtained by the Expectation-Maximization (EM) algorithm \cite{moon1996expectation}.

For the second stage of the estimation process, the copula parameters are estimated by  log likelihood maximization \cite{zhang2011efficient}.
For $d$-variate i.i.d. observations $\textbf{x} \coloneqq (\textbf{x}_1, ..., \textbf{x}_n)^t$ of size $n$ with $\textbf{x}_i \coloneqq (x_{i1}, ..., x_{id})^t$ for $i = 1, ..., n$ the corresponding log likelihood is given by
\be
l (\boldsymbol{\theta}; \textbf{x}) = \sum^{n}_{i=1} \text{log} \, c (F_1 (x_{i1}; \beta_1), ...,  
F_d (x_{id}; \beta_d); \boldsymbol{\theta}),
\ee
where $\beta _1, ..., \beta_d$ are the corresponding marginal parameters and $\boldsymbol{\theta}$ is the copula parameter space.

For model testing and verification the following goodness of fit tests are used.
The Kolmogorov-Smirnov (KS) Test is defined as the largest difference between two CDFs. The Cramer-von-Mises (CVM) test is the full sum across every observation of the difference of the two CDFs \cite{darling1957kolmogorov}. Therefore it gives a higher power gain than the KS test by using the full joint sample.

\begin{figure}
 \centering
 \includegraphics[width=0.5 \textwidth]{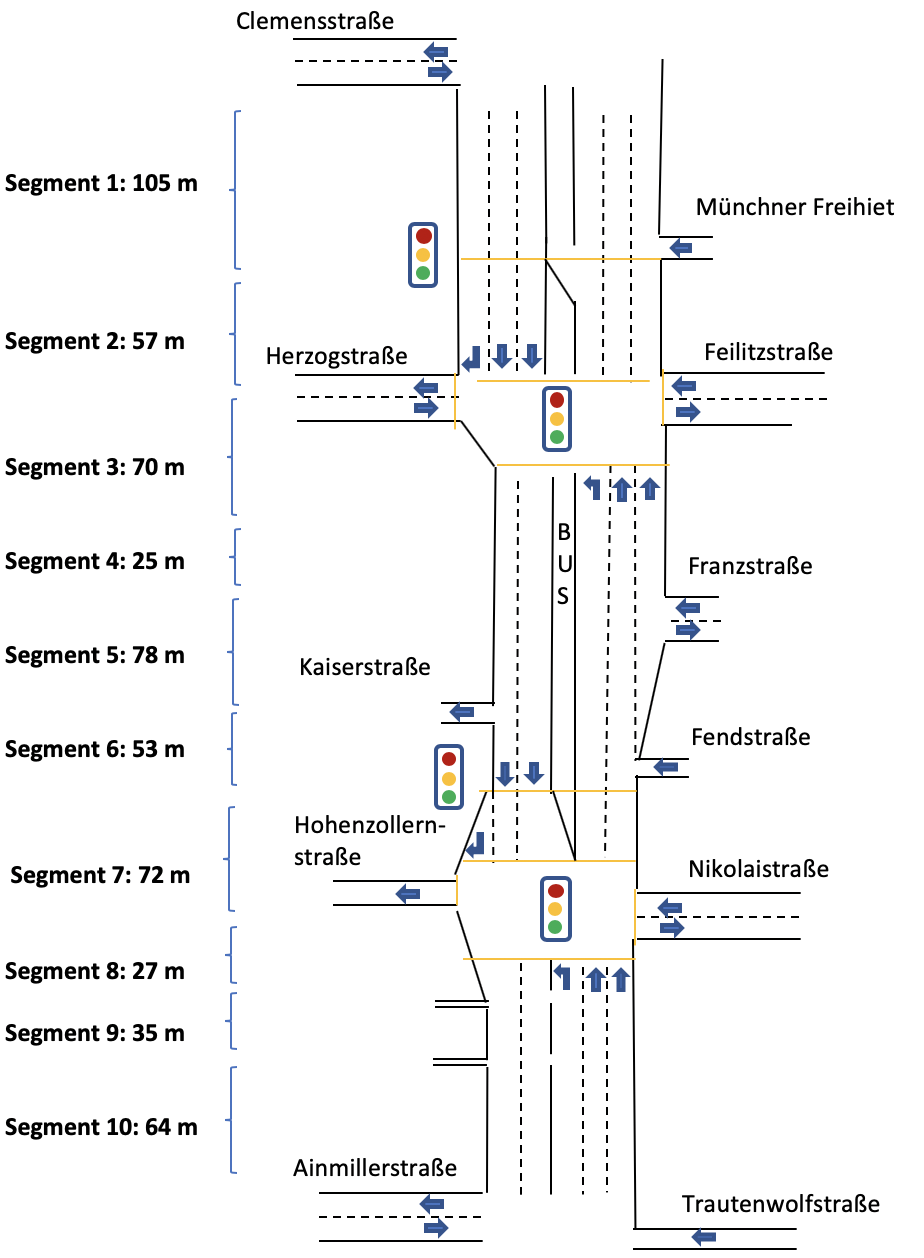}
 \caption{Study site: Major urban arterial on Leopoldstraße in Munich}
 \label{map}
\end{figure}

\section{Case Study}
Here we present the results for the case study. 
First the empirical travel time data as well as the study site are described. Then, empirical segment TTD is investigated. Correlation between successive segments is analyzed.
Finally, the results for path TTD estimation are presented. 

\subsection{Data}
The travel time data provided by BMW Group is collected from probe vehicles.
The setup includes a fleet of  probe vehicles, which have a module that reports GPS data and a central server, which collects all data in a database. Each vehicle samples the current GPS positions in intervals of 10s to 30s, which are stored in in the local memory of the vehicle together with the according timestamp. The recently sampled positions and according timestamps are transmitted to the central server. Each transmitted position is linked to an alias, which is randomly generated by the vehicle and changes over time due to protection of driver's privacy. At the server, single transmitted positions of the same alias can be connected in order to reconstruct vehicle trajectories. However since vehicles do not transmit continuously, and hide their vehicle ID, it is not possible to reconstruct complete trips or infer driver's identity.
The collected raw data is then matched to the links of the road network. Velocities are derived from the difference of time and location, respectively, between two GPS points. Travel times are then obtained using the velocity and the length of the link.

\subsection{Study site}
 
The study site consists of ten segments with a total length of 586 m on Leopoldstraße, a major urban arterial in Munich. 
A schematic illustration of the study site can be found in figure \ref{map}. The arterial comprises two signalized intersections. In addition, there is one bus lane, which stretches from Münchner Freiheit until Hohenzollernstraße, and there is signal control at the start and the end of the bus lane, respectively.  
The travel time data was collected over a period of one year, i.e. from 01. March 2013 until 01. March 2014.
For the through movement of the arterial 4495 trips were recorded. In order to obtain travel time data for the through movements of the ten segments of the arterial, the data with the same Drive-ID for each segment was chosen.

\subsection{Investigation of segment TTD}

\begin{figure}
 \centering
 \includegraphics[width=0.45 \textwidth]{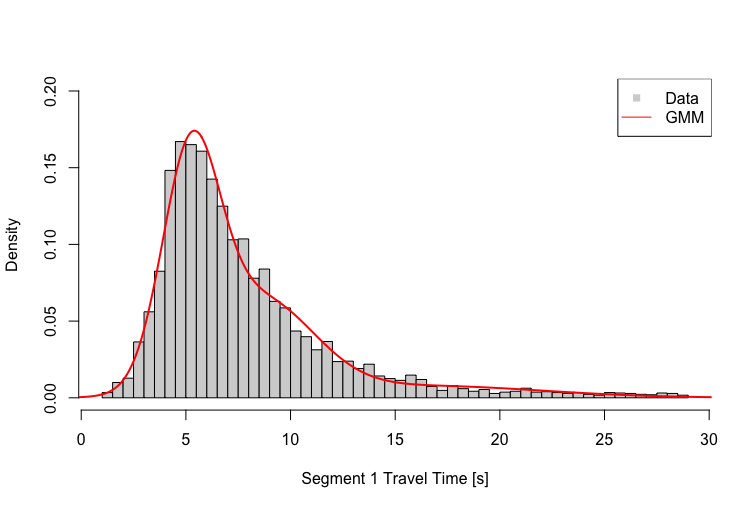}
 \caption{Segment TTD fitting for segment 2 of Leopoldstraße using GMM.  This mixture consists of three Gaussians with the following properties. The KS statistic is 0.019, which shows an accurate fit.
 The remaining segments are fitted analogously. }
 \label{GMM1}
\end{figure}

\begin{figure}
 \centering
 \includegraphics[width=0.5\textwidth]{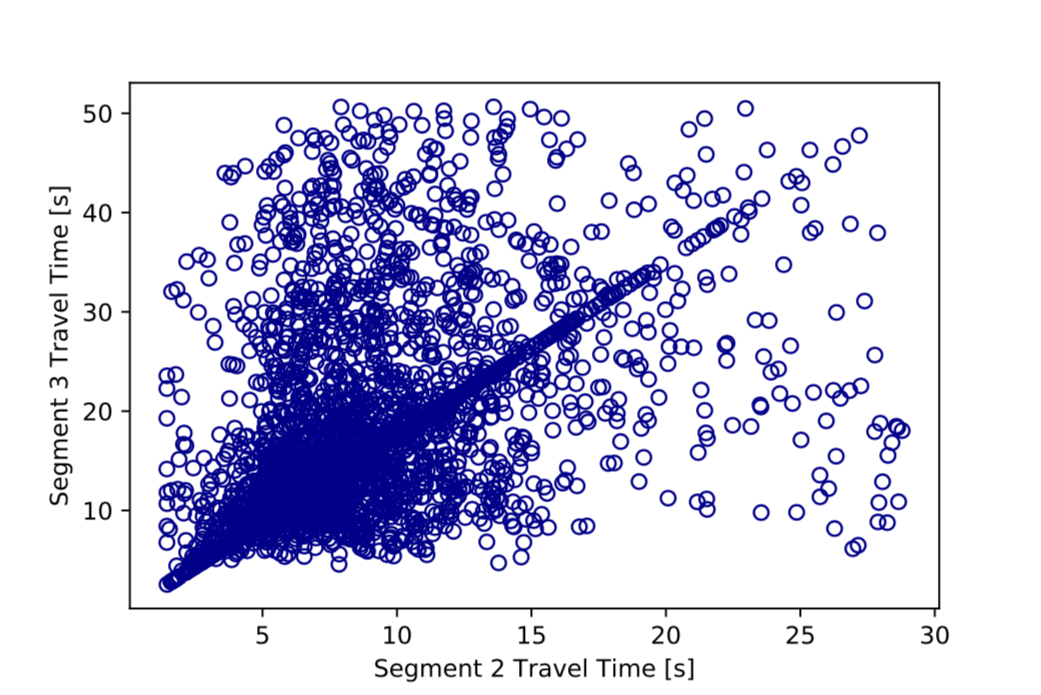}
 \caption{Empirical segment correlation analysis: a) scatterplot for the first two segments of Leopoldstraße with $\tau = 0.604$. }
 \label{cormatr}
\end{figure}

In \cite{CHEN2017} it is suggested that segment travel time on urban arterials follow multimodal distribution with three states and GMM with three components is chosen to estimate the marginal distribution. 
Also for this study, GMM with three components showed an accurate fit.
Figure \ref{GMM1} shows the TTD estimation for segment 2 and KS statistic, exemplarily. The other segments are estimated accordingly with a similar accuracy. The parameters of the GGM for each segment are listed in table \ref{gmm_parameters}.
The first component of GMM denotes the situation when vehicle travel with free flow speed. The second component implies partially delayed vehicles, while the third component denotes the congested situation.  

In order to illustrate the correlation between segments, a scatter diagram 
 for the travel times of segment 2 and segment 3 is shown in figure \ref{cormatr}, exemplarily. 
We can observe a complex correlation structure with a tendency to lower tail dependence, rather than a linear correlation. 

The line through the origin is caused by the estimation of the velocities of probe vehicles described above. If one GPS point is located in front of Segment 2 and the next GPS point is located behind Segment 3, the velocity for both segments is equal. Therefore, travel time of Segment 2 and Segment 3 are proportional, causing the line through origin, where the corresponding samples lie.

 This resembles only a small fraction of the total number of vehicles. The majority of travel times is affected by the several road features leading to an interrupted traffic flow with complex correlation structure. For that reason the rank correlation coefficient Kendall's tau is chosen as dependence measure. The values of Kendall's tau for each segment pair are listed in table \ref{taus}.

\begin{table}[h]
\caption{GMM Parameters for segment TTD estimation}
\label{gmm_parameters}
\begin{center}
\begin{tabular}{llll}
\hline
Link          & Mean ($\mu$)  &  Sigma ($\sigma$) & Weight ($\pi$) \\
\hline 
 1        &    (16.08, 31.41, 62.92)       & (5.25, 9.79, 12.65)  &  (0.31, 0.34, 0.34) \\

 2       &       (5.41, 8.86, 16.31)        & (1.44, 2.68, 5.58)   & (0.52, 0.38, 0.09) \\

 3       &     (8.92, 14.55, 29.37)         & (2.03, 4.06, 9.55)   &  (0.43, 0.34, 0.22)     \\

 4       &     (3.11, 5.72, 10.33)          &   (0.72, 1.69, 3.26)  & (0.43, 0.38, 0.17)     \\

 5       &      (9.46, 17.58, 36.01)         &   (2.18, 5.26, 10.38)  & (0.33, 0.38, 0.28)    \\

 6       &      (6.43, 12.26, 29.55)         &  (1.53, 3.94, 5.36)   & (0.46, 0.35, 0.17)    \\

 7       &       (8.24, 13.30, 27.82)        &   (1.68, 3.75, 10.94)  & (0.54, 0.37, 0.07)   \\

 8       &       (2.76, 3.97, 7.09)        &   (0.48, 0.91, 2.69)  & (0.48, 0.38, 0.12)     \\

 9       &       (3.59, 5.42, 10.17)        &  (0.64, 1.34, 3.94)   & (0.52, 0.35, 0.11)    \\

 10       &     (6.67, 11.29, 22.46)       &  (1.32, 3.22, 8.64)   & (0.52, 0.35, 0.11)     \\
\hline
\end{tabular}
\end{center}
\end{table}

\begin{table}[h]
\caption{Kendall taus for the link pairs.}
\label{taus}
\begin{center}
\begin{tabular}{ll}
\hline
Link pair          & Kendall's tau   \\
\hline 
 1, 2      &    0.318       \\

 2, 3       &     0.604         \\

 3, 4      &        0.698         \\

 4, 5       &         0.602        \\

 5, 6      &        0.417      \\

 6, 7       &       0.490       \\

 7, 8       &         0.639      \\

 8, 9       &          0.835        \\
 
 9, 10       &          0.748        \\

\hline
\end{tabular}
\end{center}
\end{table}

\subsection{Estimation of path TTD}

    \begin{figure*}[!tb]
      \centering
      \subfloat[ \label{fig:first-case1}]{%
        \includegraphics[width=0.5 \textwidth]{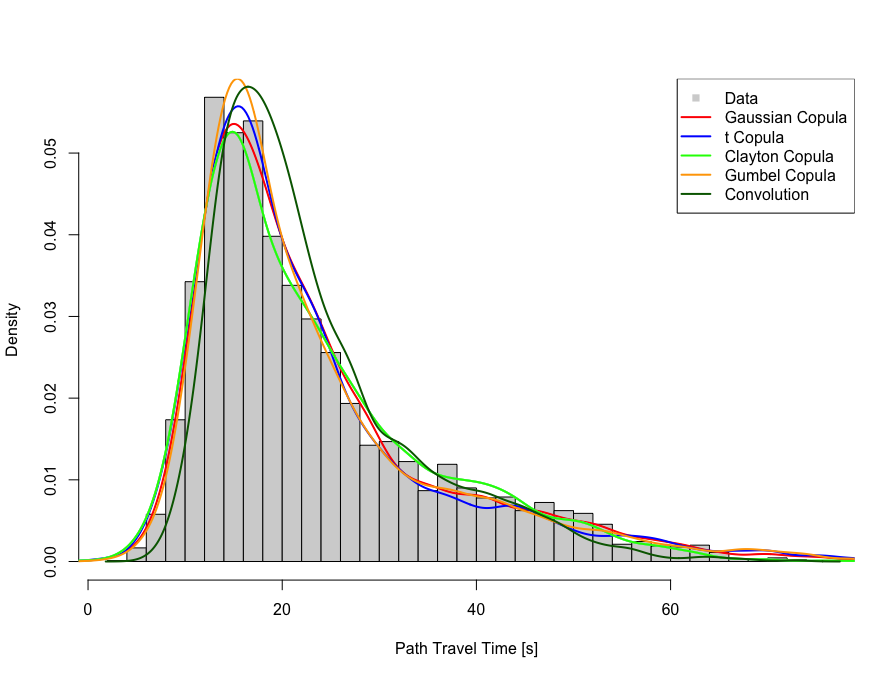}%
      }
      \subfloat[ \label{fig:second-case1}]{%
        \includegraphics[width=0.5 \textwidth]{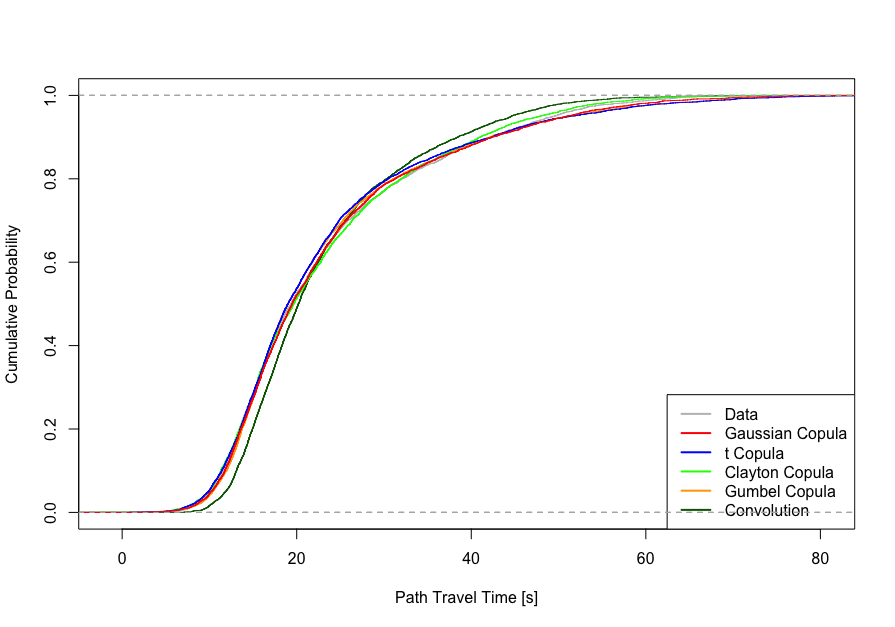}%
      }
      \caption{Path TTD estimation by 2D Models: PDF (a) and CDF (b). }
      \label{pdf_cdf_2d}
    \end{figure*}

    \begin{figure*}[!tb]
      \centering
      \subfloat[ \label{fig:first-case2}]{%
        \includegraphics[width=0.5 \textwidth]{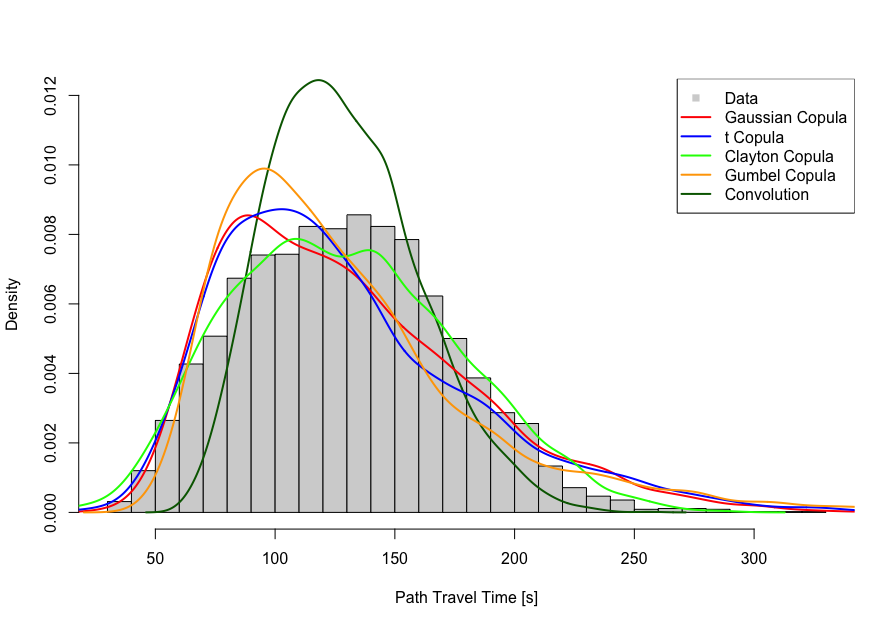}%
      }
      \subfloat[ \label{fig:second-case2}]{%
        \includegraphics[width=0.5 \textwidth]{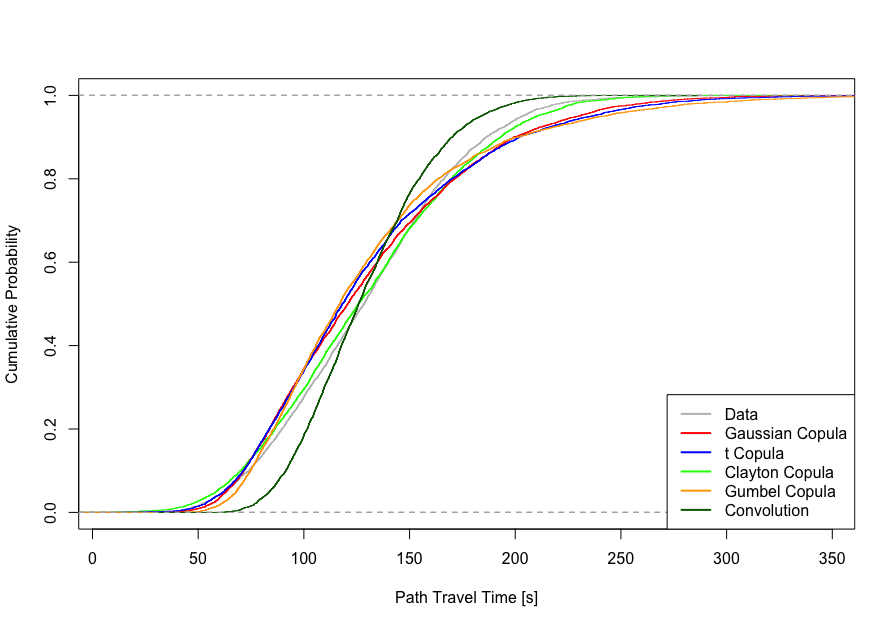}%
      }
      \caption{Path TTD estimation by 10D Models: PDF (a) and CDF (b). }
      \label{pdf_cdf_10d}
    \end{figure*}

For estimating path TTD we compare the performance of different copula models with the convolution, and the empirical distribution. 
For the copula model, the two-stage estimation procedure was used as described above. 
KS and CVM statistics are used as goodness of fit tests.  A lower value for the KS and CVM statistic, respectively, indicates a better fit.

\begin{table}[h]
\caption{Goodness of fit tests and parameters of the 2D models for path TTD estimation.}
\label{gof_urban2}
\begin{center}
\begin{tabular}{llll}
\hline
 Model         & KS   & CVM  & Parameter  \\
\hline 
2D Convolution  &  0.031   & 0.023 & /   \\

2D Gauss   & 0.015 & 0.004 &  $\rho = 0.701$ \\

2D t &  0.016  & 0.010 & $\rho=0.826$, $\nu=1.607$  \\

\textbf{2D Clayton} &  \textbf{0.014}  & \textbf{0.003} & $\alpha=2.595$ \\

2D Gumbel & 0.024  & 0.008 & $\alpha=1.993$ \\

\hline
\end{tabular}
\end{center}
\end{table}

In order to assess the scalability of the estimation models, we iteratively increase the number of aggregated segments. 
First, we estimate TTD for a path comprised of two segments and refer to the corresponding estimation models as "2D Models". Then, we estimate path TTD for the total path comprised of ten segments and refer to the corresponding models as "10D Models". 

\begin{table}[h]
\caption{Goodness of fit tests and parameters of the 10D models for path TTD estimation.}
\label{gof_urban10}
\begin{center}
\begin{tabular}{llll}
\hline
 Model                        & KS       & CVM     & Parameter  \\
\hline
10D Convolution        & 0.085   & 0.914     &  /  \\

10D Gauss.               & 0.046    & 0.340     & $\rho =0.387$ \\

10D t                         & 0.053      & 0.472      & $\rho=0.428$, $\nu=6.582$ \\

\textbf{10D Clayton} & \textbf{0.026}   & \textbf{0.061} & $\alpha=0.698$ \\

10D Gumbel & 0.054   & 0.579  & $\alpha=1.363$ \\
\hline
\end{tabular}
\end{center}
\end{table}

For evaluating the performance of the 2D models, we estimated TTD for a path comprised of Segment 2 and Segment 3. Figure \ref{pdf_cdf_2d} shows the corresponding PDF and CDF. Goodness of fit tests and the parameters of each copula model are listed in table \ref{gof_urban2}. Each copula model performs better than the convolution due to their ability to incorporate segment correlation. The Clayton copula performs best. A possible reason may be its ability to capture lower tail dependence.

PDF and CDF for TTD estimation by the 10D models for the total path comprised of ten segments is shown in figure \ref{pdf_cdf_10d}. The corresponding goodness of fit tests and the parameters of each copula model are listed in table \ref{gof_urban10}. Compared to the results for the 2D models, the inaccurate estimation of the convolution as well as the superior estimation of the copula models is more distinct. Again, each copula model performs better than the convolution, while the Clayton copula shows the best fit. 

\begin{figure}
 \centering
 \includegraphics[width=0.5 \textwidth]{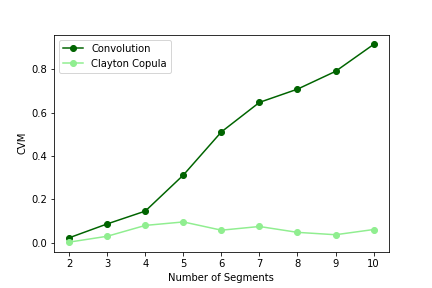}
 \caption{CVM statistic for path TTD estimation by the convolution and Clayton copula for the iterative aggregation of the ten segments.}
 \label{cvm}
\end{figure}

Figure \ref{cvm} shows the CVM statistic for path TDD estimation by the convolution and the Clayton copula for the iterative aggregation of the ten segments. The accuracy of the convolution decreases with the number of aggregated segments, while the accuracy of the Clayton copula stays nearly constant.
Therefore, using convolution for estimating path TTD will lead to severe inaccuracies as successive segments are treated as independent. This makes convolution ineligible for assessing travel time reliability. Copulas are able to incorporate segment correlation and, thus, provide an accurate assessment of travel time reliability.

\section{Conclusion and Outlook}
This paper presented a copula-based approach to aggregate individual segment TTDs to estimate path TDDs. The aim was to evaluate the scalability of different copula models in terms of number of aggregated segments with real day-to-day data.

GPS travel time data was collected from probe vehicles over the period of one year for a major urban arterial in Munich, Germany.
First segment TTDs were investigated. Marginal distributions were estimated using GMM with three components. Segment correlation was analyzed showing a lower tail dependence between successive segments.
Path TTD estimation models were first assessed for a path comprised of two segments, then for a path comprised of ten segments. 
Gaussian, Student-t, Clayton, and Gumbel Copulas were compared to the empirical path TTD and the convolution.
 The main findings are the following: 
 \begin{enumerate}
 \item Path TTD estimation by each copula model is more accurate than the estimation by the convolution. 
 \item The copula model has potential to model path TTD for an increasing number of segments, whereas the accuracy of the convolution decreases with the number of aggregated segments.
 \item The Clayton copula is able to incorporate the segment correlation most accurately compared to other copula models. A possible reason may be its ability to capture lower tail dependence.
 \end{enumerate}
 
Future work will focus on further investigating segment correlation by clustering day-to-day data dependent on time of day and day of week.
In addition, different study sites will be investigated to improve the applicability of the proposed methodology in field implementation.

\bibliographystyle{plain}
\bibliography{proceedings}

\end{document}